\documentclass{aa}  
\usepackage{graphicx}
\usepackage{txfonts}
\begin{document}
   \title{On the fidelity of the core mass functions derived from dust column density data}


   \author{J. Kainulainen \inst{1 \and 2} 
          \and 
          C. J. Lada \inst{3}
          \and 
          J. M. Rathborne \inst{3}
          \and
	  J. F. Alves \inst{4}
          }

   \offprints{jouni.kainulainen@helsinki.fi}

   \institute{TKK/Mets\"ahovi Radio Observatory, Mets\"ahovintie 114, FIN-02540 Kylm\"al\"a, Finland \\
              \email{jouni.kainulainen@helsinki.fi}
         \and Observatory, P.O. Box 14, FIN-00014 University of Helsinki, Finland
         \and Harvard-Smithsonian Center for Astrophysics, Mail Stop 72, 60 Garden Street, Cambridge, MA 02138, USA
         \and Calar Alto Observatory, Centro Astronomico Hispano, Alem\'an, C/q Jes\'us Durb\'an Rem\'on 2-2, 04004 Almeria, Spain
  }

   \date{Received <>; accepted <>}

 
  \abstract
   {}
   {We examine the recoverability and completeness limits of the dense core mass functions (CMFs) derived for a molecular cloud using extinction data and a core identification scheme based on two-dimensional thresholding. We study how the selection of core extraction parameters affects the accuracy and completeness limit of the derived CMF and the core masses, and also how accurately the CMF can be derived in varying core crowding conditions.}
   {We performed simulations where a population of artificial cores was embedded into the variable background extinction field of the Pipe nebula. We extracted the cores from the simulated extinction maps, constructed the CMFs, and compared them to the input CMFs. The simulations were repeated using a variety of extraction parameters and several core populations with differing input mass functions and differing degrees of crowding.}
   {The fidelity of the observed CMF depends on the parameters selected for the core extraction algorithm for our background. More importantly, it depends on how crowded the core population is. We find that the observed CMF recovers the true CMF reliably when the mean separation of cores is larger than the mean diameter of the cores ($f>1$). If this condition holds, the derived CMF for the Pipe nebula background is accurate and complete above $M\gtrsim 0.8\dots 1.5$ M$_\odot$, depending on the parameters used for the core extraction. In the simulations, the best fidelity was achieved with the detection threshold of 1 or 2 times the rms-noise of the extinction data, and with the contour level spacings of $3$ times the rms-noise. Choosing larger threshold and wider level spacings increases the limiting mass. The simulations also show that when $f \gtrsim 1.5$, the masses of individual cores are recovered with a typical uncertainty of $25\dots 30$ \%. When $f\approx 1$ the uncertainty is $\sim 60$ \%. In very crowded cases where $f< 1$ the core identification algorithm is unable to recover the masses of the cores adequately, and the derived CMF is unlikely to represent the underlying CMF. For the cores of the Pipe nebula $f \approx 2.0$ and therefore the use of the method in that region is justified.}
    {}
   \keywords{dust, extinction $-$ ISM: clouds $-$ ISM: structure $-$ stars: formation $-$ stars: luminosity function, mass function}

   \maketitle
%

\section{Introduction}

The dense cores of interstellar molecular clouds are the precursors of stars, providing the suitable physical conditions and the mass reservoir for the star-forming process to ensue. Recently, several studies of the dust emission from such cores have presented intriguing observational evidence suggesting that the mass function of the dense cores (CMF) could be directly linked to the initial mass function of stars (IMF). In particular, the CMFs derived for nearby molecular clouds  have been found to resemble the high-mass, power-law slope of the stellar IMF presented by Salpeter (\cite{salpeter55}) (Motte et al. \cite{motte98}; Testi \& Sargent \cite{testi98};  Johnstone \cite{johnstone00}, \cite{johnstone01}; Motte et al. \cite{motte01}; Johnstone \& Bally \cite{johnstonebally06}; Johnstone et al. \cite{johnstone06}; Stanke et al. \cite{stanke06}; Reid \& Wilson \cite{reid06}a,b; Young \cite{young06}; Nutter \& Ward-Thompson \cite{nutter07}; Enoch et al. \cite{enoch08}; Simpson et al. \cite{simpson08}). Even though the uncertainties in the slopes of the derived CMFs are often large, the slopes seem to be in agreement with that of the stellar IMF for stellar masses greater than roughly 0.5 M$_\odot$. Because of the scale-free nature of power-laws, it is difficult to ascertain, from the similarity of these slopes alone, the true nature of the physical connection between the CMF and stellar IMF. However, in some of these studies, the derived CMF is also interpreted to flatten or to turn over at lower masses, making the similarity between the CMF and IMF even more striking and providing a physical scaling or characteristic mass for the CMF which can be then compared to the characteristic mass similarly measured for the IMF. However, there is considerable variations in the possible peak position of the CMF among the studies, with the estimates ranging from $\sim 0.1$ M$_\odot$ to $\sim 8$ M$_\odot$, while in comparison, the charasteristic mass scale for the IMF is relatively well-determined to be $\sim 0.5$ M$_\odot$ (Kroupa \cite{kroupa02}).

It is quite possible that a large part of the observed variation between these CMFs results from the uncertainties inherent in their construction. Deriving the CMF accurately for a molecular cloud is not trivial because of  three main difficulties. First, gathering the sensitive, uniformly calibrated data required to compile a sufficiently large sample of cores is a challenge for any current observational technique. The variations in the shapes of the derived CMFs that are reported in the studies mentioned above are likely, in part, due to sampling errors resulting from the small populations considered (typically 20 \dots 100 cores). Second, deriving a robust CMF requires an accurate and unbiased determination of individual core masses over a wide mass range. This, in turn, requires an accurate observational techinque for measuring core mass. However, it is often difficult to assess the uncertainties inherent in a given methodology for measuring mass and consequently the effects of these uncertainties on the derived CMF. 
Third, identifying cores within a cloud requires a physically meaningful and clear operational definition of a dense core. Unfortunately, there is no concensus among observers on what exactly consititues such a definition and different definitions are usually employed in the various studies.
It is not simple to assess the effect of these differing definitions on the derived CMFs. 

In this paper, our goal is to determine the reliability of the CMFs and the core masses derived from 2-dimensional column density maps. For this purpose we have conducted simulations where an artificial population of dense cores is embedded into an observed (real) field of variable dust column density. The cores are then extracted using a specific core extraction (definition) scheme and a CMF constructed from them. This is then compared  to the original, input CMF.

In order to minimize the difficulties discussed above, we have chosen to model dust extinction observations in these numerical experiments because, in contrast to the observations of dust emission, interpreting extinction data in terms of dust column density (and core mass) does not depend on the temperature of dust grains and only very weakly on the optical properties of the grains (i.e. the extinction law), which are well known and calibrated in the near-infrared (e.g. Mathis \cite{mathis90}). In particular, we have chosen to simulate the dust column density conditions in the Pipe Nebula, a nearby cloud with well-populated and smooth distribution of background stars. These properties have enabled  wide-field, near-infrared extinction mapping of the cloud (Lombardi et al. \cite{lombardi06}, LAL06 hereafter) which provides both a good sensitivity to the low contrast features and a high spatial resolution which are required to accurately measure the masses of cores over a wide mass range.

Recently, Alves et al. (\cite{alves07}, ALL07 hereafter) presented a study of the mass spectrum of the dense cores in the Pipe nebula. In their work, ALL07 used the extinction map of LAL06 to trace the dust column density. With a wavelet decomposition technique and an automated thresholding routine, \textsf{clfind2d}, they identified 159 dense core, measured their masses and constructed the CMF for the Pipe nebula. Consistent with the previous dust emission studies, ALL07 found that the CMF is surprisingly similar in shape to the stellar IMF, including a break or flattening at low mass. The break they measured in the Pipe nebula CMF suggested a characteristic core mass of $M=2\dots 3$  M$_\odot$, higher than the corresponding value of the stellar IMF. This observation supported the view that the IMF might originate directly from the dense core mass function after modification by a uniform star formation efficiency.

To simulate various populations of dense cores in a realistic extinction field, we create simulated extinction images of the Pipe Nebula by embedding populations of artificial dense cores into the real, variable large-scale extinction component of the Pipe Nebula. We then analyze these simulated images in the same way as the observers (ALL07) and then evaluate the fidelity of the extracted CMFs. The details of the simulations are given in Section \S\ref{sec:simulations}, and the results are presented and discussed in Sections \S\ref{sec:results} and \S\ref{sec:discussion}. In Section \S\ref{sec:conclusions} we give our conclusions. 

\section{Simulations}
\label{sec:simulations}

We simulated the process of deriving the CMF by extracting cores from an  artificial extinction image. The core extraction was repeated using several sets of input parameters for the identification algorithm. The main characteristics of the core population, namely the form of the core mass function and the degree of crowding, were also altered to study their effect on the observed CMF. We made numerous realizations of each simulation to lower the counting errors of the histogram bins of the observed CMF, and thus to better isolate the effect of different variables. In the following the construction of the simulated extinction maps and the core extraction procedure are explained in more detail.

\subsection{Simulated extinction maps}
\label{sec:simulated_extinction_maps}


One of the main difficulties in constructing the CMF for a molecular cloud from extinction data is disentangling the dense cores from a variable "background" extinction caused by the more diffuse dust component of the cloud. In order to simulate the effects of the variable and noisy background in a realistic manner, we used real extinction data to create the background component for the simulations. We exploited the extinction map of the \object{Pipe nebula}, derived from near-infrared color-excess data by LAL06, to create the map of the diffuse dust component. This $8^\circ \times 6^\circ$ wide extinction map is presented in Fig. 7 of LAL06, and it features a dynamical range of $A_\mathrm{V}=0\dots 25$ mags with the 3-$\sigma$ error of 0.5 mags at the FWHM resolution of $1'$. At the distance of 130 pc (LAL06) the dimensions of the map translate to $\sim 18 \times 14$ pc, and the resolution to 0.038 pc.

We extracted the dense cores initially present in the extinction map by using the wavelet decomposition technique described in ALL07. We used the technique to produce a map of extended structures (the "cores-only" map) at $4'$ scale and subtracted it from the original extinction map. The remaining, core subtracted map was used as the large scale background extinction component in our simulations. Thus, the simulated extinction maps have the same angular extent and the same resolution as the extinction map presented by LAL06, but prominent structures at angular scales smaller than 4' ($\sim 0.15$ pc) have been largely filtered out from it. The range of extinction values in this background map is $A_\mathrm{V}=0\dots 10$ mags, and the frequency distribution of the values is similar to that of the original extinction map when $A_\mathrm{V}<8$ mags (see Fig. 20 of LAL06). At higher extinctions the frequency distribution of the background map goes quickly to zero. The rms-variation in the background map is a function of position, but a typical value for it is $\sigma^\mathrm{rms}\approx 0.4$ mags.


To mimic the population of dense cores with non-constant column density profiles in the cloud, we embedded an assemblage of elliptical Gaussians into the map of the diffuse background extinction. The individual Gaussians, i.e. the cores, were created in the following way. First, the mass of a core was drawn randomly from the input mass function. These input mass functions were defined using power-law distributions\footnote{We note that the mass function is calculated per unit log mass. In this notation the Salpeter (\cite{salpeter55}) mass function has the exponent $\Gamma=-1.35$.}, i.e. $dN/d \log M = M^{\Gamma}$. In particular, we used $\Gamma=-1$, $\Gamma=0$, and a broken power-law where $\Gamma=-0.3$ ($M < 3 $ M$_\odot$), and $\Gamma=-1.35$ ($M > 3 $ M$_\odot$). The range of the input mass function was always set to $\log M=-0.55 \dots 1.2$. Second, a radius was calculated for the core from the mass-to-radius relation $\log{2R} = 0.35 \times \log M + 0.65$ that was derived for the cores of the Pipe nebula by Lada et al. (\cite{lada08}). In the equation $R$ is given in arcminutes and $M$ in solar masses. The total mass and radius were then used to calculate the dispersion ($\sigma$) of the Gaussian profile. The mass of a core was distributed to a region within 3$\sigma$ distance from its center point using the profile as a weighting function. The resulting sizes of the cores range between $\sim1.7'$-$6.5'$ (0.06-0.25 pc), and the upper end of their size range overlaps with the lower end of the range of the structures in the background map. Finally, both a uniformly distributed ellipticity (0.5-1) and position angle (0-2$\pi$) were applied to the core. The resulting column density distribution of the core was embedded into the background extinction map with a simple summation. In total, about 250 cores were embedded to the background in each realization of the simulations.


The spatial distribution of the cores is one of the main parameters whose effect we want to study in this work. This is because the 2-dimensional core extraction schemes, such as \textsf{clfind2d} used by ALL07 and this study (see Section \ref{sec:core_extraction}), can not easily disentangle overlapping cores, and it is likely that the derivation of CMF is severely hampered in very crowded regions. Obviously, the crowding of the cores depends not only on their separation, but also on their size. To take both of these parameters into account, we used the ratio $f$ = mean core separation / mean core diameter as a metric to describe the crowding of cores in the simulations. By separation we refer to the distance measured from a core to its nearest neighbor.

The positions of the cores in the simulations were chosen randomly from the simulated image. The randomization was weighted with the spatial density function of the real cores identified from the Pipe nebula by ALL07. The spatial distribution of the cores in the Pipe nebula is characterized by the mean nearest neighbor distance of $9.5'$, which together with the mean core size of $4.7'$ gives the ratio $f=2$. If the medians are used instead of means, the ratio $f=1.7$ follows. The spatial distribution of the simulated cores can be tuned to have different ratios of $f$ by choosing the size of the kernel function that is used to calculate the spatial density function of the cores. In this way, we generated distributions where $f=2$, 1.5, 1, and 0.5. These cover the range where we expect both minor ($f=2$) and substantial ($f=0.5$) degrees of crowding. 

\begin{figure*}
   \centering
   \includegraphics[width=1.99\columnwidth]{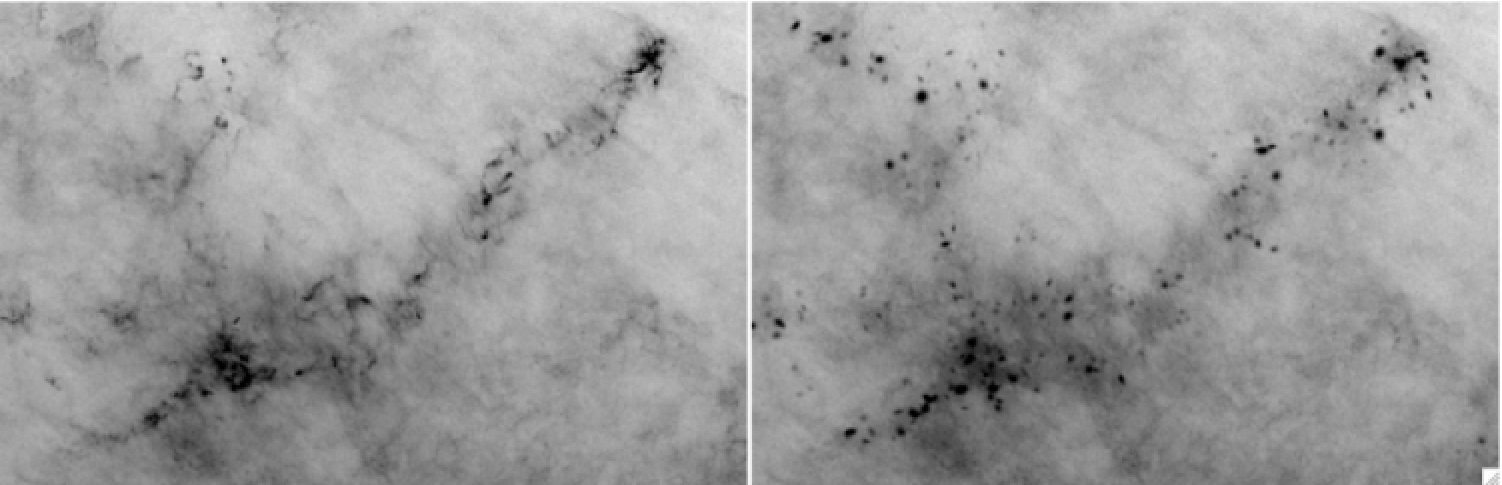}
   \includegraphics[width=1.99\columnwidth]{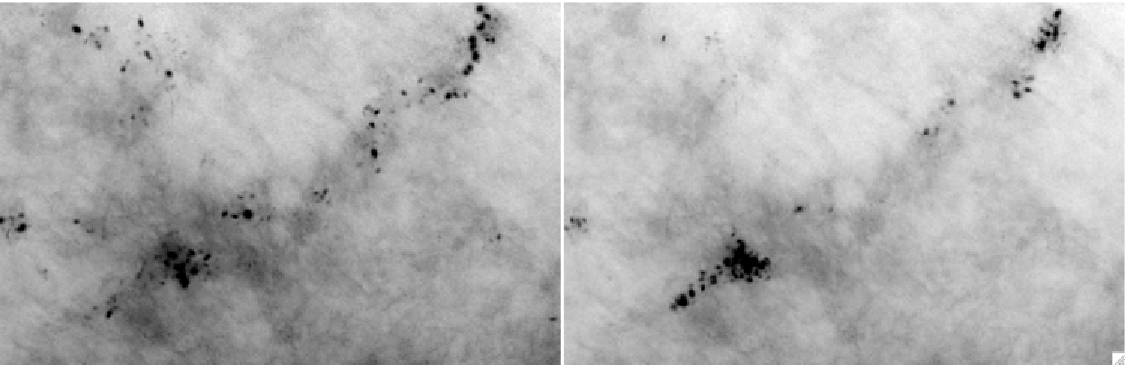}
      \caption{Illustration of the simulated extinction data. {\bf Top Left: } The extinction map of the Pipe nebula, as presented in LAL06. The mean separation to mean diameter ratio of the cores detected by ALL07 is $f=2.0$ {\bf Top Right: }An example of the extinction map used in the simulations. The map is composed of a diffuse background component similar to the large-scale structure of the Pipe nebula, and of a population of artificial cores (elliptical Gaussians). The figure corresponds to the simulation where $\Gamma =-1$ and $f=1.5$. {\bf Bottom Left: } The same, but for simulation where $f=1.0$. {\bf Bottom Right: }The same, but for simulation where $f=0.5$. 
                    }
         \label{fig:simulated-cores}
   \end{figure*}

An illustration of the simulated extinction maps is shown in Fig. \ref{fig:simulated-cores}. The figure shows the actual extinction map of the Pipe nebula, and the extinction maps for three simulations ($f=1.5$, 1.0, and 0.5). Each simulated extinction map is a sum of the variable background component and the population of embedded cores. On a large scale these maps are comparable to the original extinction map of LAL06.

\subsection{The core extraction}
\label{sec:core_extraction}

We use a two-step procedure to construct the CMFs from the simulated extinction maps. The procedure is the same as used by ALL07, and it is outlined in their paper (see also \S2.1 of Lada et al. \cite{lada08}). First, the extinction map was decomposed with a wavelet transformation at 4$'$ scale. The wavelet transform effectively removes the variations at scales larger than the selected spatial scale. This results in a "cores only" map that includes only the small-scale structure of the data. Second, the cores were identified from the cores only map using the routine \textsf{clfind2d} (a 2D version of the routine \textsf{clumpfind} presented by Williams et al. \cite{williams94}). The routine identifies the cores based on 2-dimensional contour line cutting. The input parameters for the routine are the contour levels used in the identification, the lowest of which sets the threshold level for detection. Our main goal in this work is to examine how the selection of parameters for \textsf{clfind2d} affects the reliability and the completeness limit of the observed CMF, and thus we performed the extraction using ten different sets of parameters. In these sets we combined both low and high thresholds, and narrow and wide level spacings. The contour levels are listed in Table \ref{tab:parameters} and are given as multiplicatives of $\sigma^\mathrm{rms}$, which is the rms-variation in extinction values within the background map ($\sigma^\mathrm{rms}=0.4$ mags). We note that ALL07 used contour levels $[1.2, 4, 6]$ mags that are close to the lowest levels of the parameter set \#3. They also required a "visual verification" of the cores to be included in the CMF.

We ran the \textsf{clfind2d} routine on the simulated extinction maps using all contour levels listed in Table \ref{tab:parameters}. The routine outputs the characteristics of each identified core, including peak $A_\mathrm{V}$ and its position, total $A_\mathrm{V}$, and the pixels belonging to the core. In this paper, when we refer to "an observed core" we refer to all the pixels assigned to a core by \textsf{clfind2d}. The masses of the cores were calculated from the total $A_\mathrm{V}$ within each core using equation $M = 0.0074 \times A_\mathrm{V}^\mathrm{tot}$ M$_\odot$ mag$^{-1}$. This equation assumes the standard gas-to-dust ratio of $N(\mathrm{H}) / A_\mathrm{V} = 2 \times 10^{21}$ cm$^{-2}$ mag$^{-1}$ (Bohlin et al. \cite{bohlin78}), the mean molecular mass of 2.35, and the distance of 130 pc to the Pipe nebula (LAL06). The mass functions were then constructed for each set of contour levels separately. We used the equation $N_\mathrm{bins} = \Delta \log M / (2 \times n^{1/3})$ for the histogram binning, where $\Delta \log M$ is the mass range and $n$ is the total number of cores.

\begin{table}
\begin{minipage}[t]{\columnwidth}
\caption{The contour levels used in the core extraction.} 
\label{tab:parameters} 
\centering
\renewcommand{\footnoterule}{}  
\begin{tabular}{c c c}
\hline\hline      
\# & $A_\mathrm{V}^{th}$ [$\sigma^\mathrm{rms}$]\footnote{The detection threshold of the core extraction algorithm} & $\delta A_\mathrm{V}$ [$\sigma^\mathrm{rms}$]\footnote{The contour level spacing} \\ 
\hline                      
   1 & 7 & 7 \\
   2 & 7 & 3  \\  
   3 & 3   & 7\\  
   4 & 3   & 5 \\ 
   5 & 3   & 3 \\   
   6 & 3   & 1$^c$ \\
   7 & 2 & 3 \\
   8 & 2 & 1\footnote{Contour increment $8 \sigma^\mathrm{rms}$ when $A_\mathrm{V}>5$ mags} \\ 
   9 & 1 & 3 \\
   10 & 1 & 1$^c$ \\ 
\hline                  
\end{tabular}
\end{minipage}
\end{table} 

\section{Results} 
\label{sec:results}

\subsection{The detectability and masses of cores}
\label{sec:detectability}


Our first goal is to study how well the cores can be identified from the variable background extinction using the method adopted in this paper. To do this, the crowding of cores needs to be eliminated from the simulations. For completely isolated cores, possible inaccuracies in the derived CMF result from the inability of the method to disentangle the cores from the extended, diffuse background. To quantify this effect, we performed a series of simulations where the cores were positioned into a fixed grid pattern so they were isolated and did not overlap. In every other respect the simulations were performed as explained in Section \S\ref{sec:simulated_extinction_maps}. We also studied how the choice of contour levels for the core extraction algorithm affects the appearance of the CMFs in these simulations.

\subsubsection{Fidelity and completeness limits}

It should be noted that it is not trivial to present a metric that would quantify the completeness of the observed CMFs in a generally meaningful way. Ideally, the completeness at each input mass could be determined from the simulations by matching every input core with an output core (or with a non-detection). Then, the convolving function which transforms the input masses to output masses could be determined. However, this can only be done if every input core can be uniquely related with an output core (or with a non-detection). In a crowded case where several cores are blended together, and are recognized as one entity by the core extraction algorithm, this will not be the case. 

In this work we make an effort to define practical mass limits, above which the observed CMFs can be regarded as good reproductions of the true CMFs. In practice, we do this simply by comparing the mass bins and the overall shape of the input and output CMFs. We will define these mass limits as the mass bins below which less than 90 \% of the cores are detected, and we will refer to these limits as \emph{fidelity limits}. We note that in a particular mass bin, having a lower number of cores than the input is not necessarily due to cores that are not detected, but may also be due to the cores whose masses are determined inaccurately. To provide insight on this, we also inspected how the observed masses of individual cores correspond to their input masses (Section \S\ref{sec:Min_vs_Mout}). 

\begin{figure*}
   \centering
   \includegraphics[width=1.5\columnwidth]{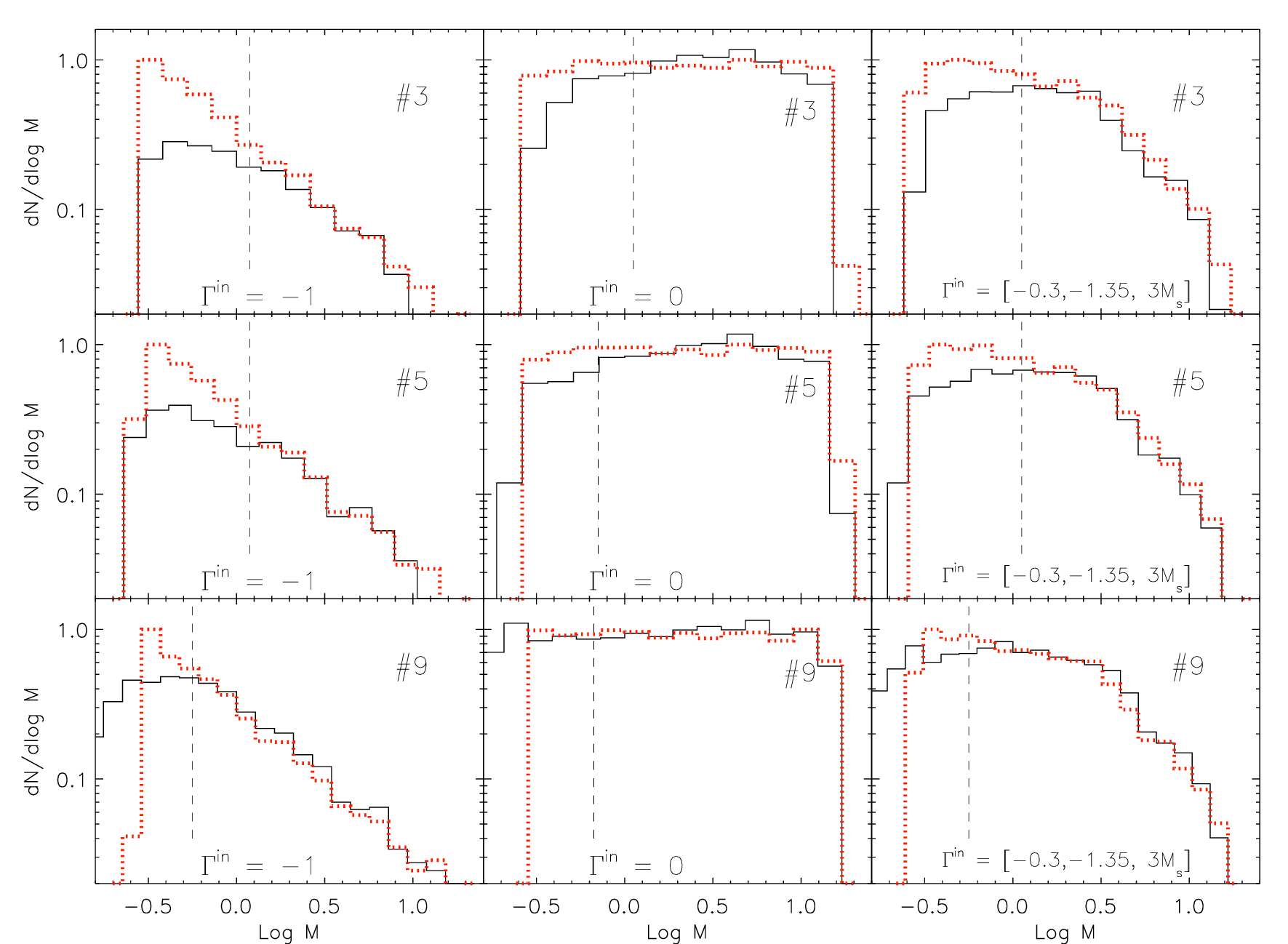}
      \caption{The CMFs derived for three simulations where the input CMF was different. The input CMFs were $\Gamma=-1$ (leftmost column), $\Gamma=0$ (center column), and $\Gamma=[-0.3,-1.35]$ with a break point at 3 M$_\odot$ (rightmost column). In these simulations the cores were positioned in an equally spaced grid pattern, so that they are isolated and do not overlap. The observed CMF is plotted with a solid line, and the input CMF with a dotted red line. The dashed vertical lines show the mass below which the derived CMF less than 90 \% complete. The CMFs derived with three different core extraction parameter sets are shown: \#3 (top row), \#5 (middle row), and \#9 (bottom row). The histograms are normalized to the peak value of the input CMF. The range of input masses was $\log M = -0.55 \dots 1.2$. The input CMFs are always the same in each panel of the columns. However, the histograms are binned equally with the observed CMFs, and may therefore slightly differ from each other.  
                    }
         \label{fig:mfs_set_and_gamma}
   \end{figure*}

Figure \ref{fig:mfs_set_and_gamma} shows the CMFs resulting from the simulations where the cores do not overlap. The three columns of the figure correspond to the simulations where the exponents of the input CMFs were $\Gamma=-1$, $\Gamma=0$, and $\Gamma=[-0.3, -1.35]$ with the breakpoint at 3 M$_\odot$. The three rows show the CMFs derived using contour levels \#3, \#5, and \#9 (see table \ref{tab:parameters}). We use these three contour levels in our illustrations throughout the paper, because they form a representative selection where we can possibly see the effects of the contour level parameters, $A_\mathrm{V}^{th}$ and $\delta A_\mathrm{V}$, on the derived CMF separately. 
The histograms of the input CMFs are also plotted in the panels. The fidelity limits, as defined earlier, are marked in the figure with dashed vertical lines. Generally, the simulations recover the shape of the input mass function well above  the fidelity limits, while below them the observed CMF becomes flatter than the input CMF. The fidelity limits are not affected by crowding in this particular simulation, and thus they indicate the mass range over which the method is able to reproduce the underlying CMF for isolated cores accurately.

The fidelity limits of the simulations in this experiment vary from $M\approx 0.6$ M$_\odot$ for sets \#7-\#10 to $M\approx 1.2$ M$_\odot$ for sets \#1-\#6. Thus, it appears that the CMF is recovered over a larger range when choosing lower thresholds and narrower level spacings for the extraction algorithm. However, it is noteworthy that there is only very subtle differences in the observed CMFs in the mass range where the CMF is recovered well with all parameter sets, i.e. at $M>1.2$ M$_\odot$. This suggests that for isolated cores above that mass the derived CMF is insensitive to the exact choice of extraction parameters. Above the fidelity limits the exponents of the CMFs are also accurately recovered, which was verified by fitting a power-law to the observed mass bins. In the simulations where the input CMF featured a break, the position of the observed break was also consistent with the input.

In the above, we defined the fidelity limit as a mass above which the derived CMF is an accurate reproduction of the input CMF. It should be noted that below this limit the detection efficiency of the cores does not necessarily drop immediately. Instead, the shape of the CMF may be altered due to the cores whose masses are determined incorrectly because of blending with the background or because of the inaccurate performance of the core extraction algorithm. Therefore, we also define actual \emph{completeness limits} which reflect the mass below which the detection efficiency of the method decreases significantly.

In the simulations where the cores do not overlap, it is possible to match each input core uniquely with an observed core or with a non-detection. We did this by comparing the center position of each input core with the observed cores of that simulation. If there was a pixel of any observed core at the center position, the input core was counted to be detected. Performing this check for all input cores of a simulation enables the examination of completeness, i.e. the ratio of the number of detected cores  and the total number of cores, as a function of the input core mass. The 90 \% completeness limits resulting from this calculation were between $M=0.45$-$0.7$ M$_\odot$, and particularly $M=0.7, 0.65$, and 0.45 M$_\odot$ for the contour levels \#3, \#5, and \#9, respectively. Below these masses, the detection efficiency decreases rapidly.

It is possible to assess the decrease of the detection efficiency caused by the wavelet filter by performing the same completeness calculation with a very low detection threshold of the \textsf{clfind2d} algorithm (e.g. 0.1 mags). In that case, practically all structures coming through the wavelet filter are broken into cores. Thus, if there is no observed core at the center position of an input core, it can only be because the core has not passed the wavelet filter and does not appear in the cores only map at all. The mass below which the detection efficiency resulting from this calculation goes below 90 \%  is $M\approx 0.4$ M$_\odot$. Thus, the detection efficiency is dominated by the properties of the contouring algorithm instead of those of the wavelet filter.

\subsubsection{The accuracy of detected masses}
\label{sec:Min_vs_Mout}

The simple comparison of the derived CMFs with the input CMFs does not provide information on whether, or how, the masses of individual cores are altered in the extraction process. It was also noted in the previous section that the direct comparison does not necessarily provide meaningful fidelity limits for crowded simulations. To have another kind of metric for the feasibility of the extraction method, we calculated the ratio of observed and input masses, $M_\mathrm{observed} / M_\mathrm{input}$, for each input core. 

\begin{figure}
   \centering
   \includegraphics[width=\columnwidth]{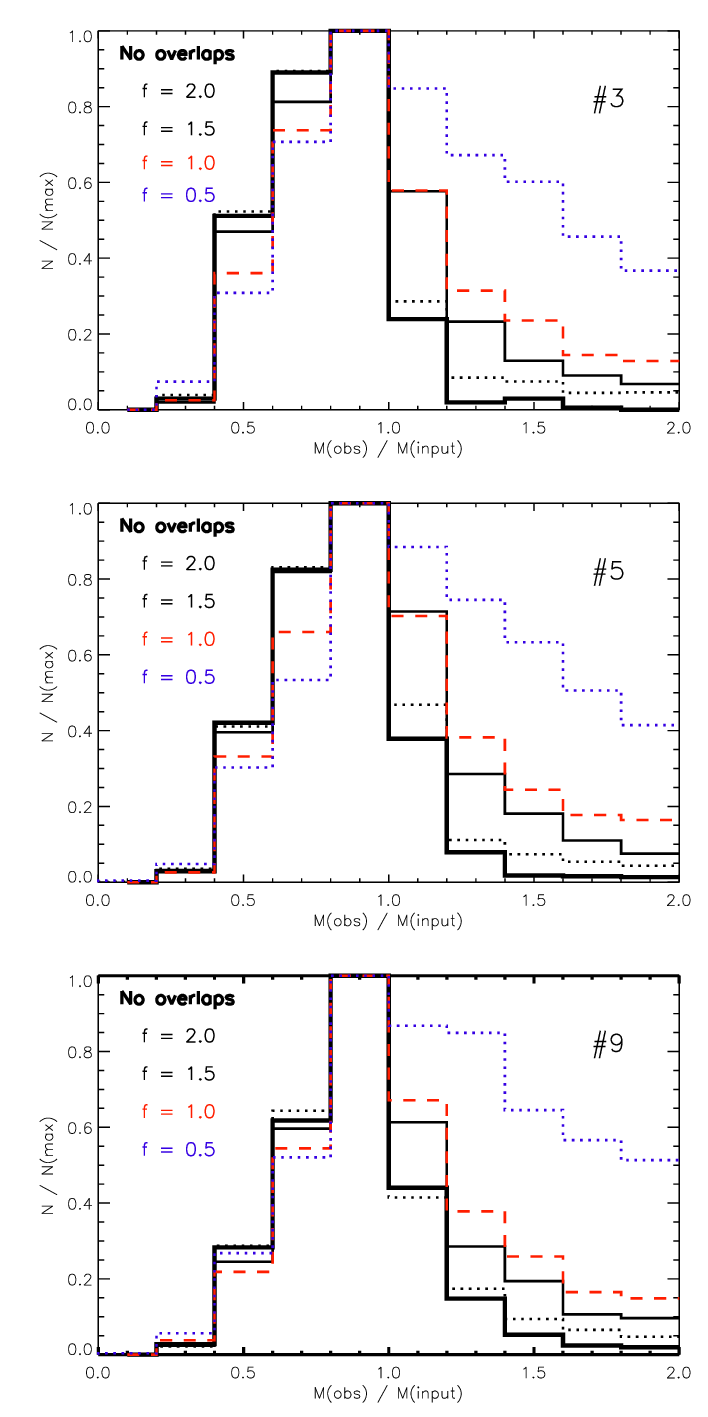}
      \caption{Frequency distributions of the $M^\mathrm{observed}/M^\mathrm{input}$ ratio for the simulations where $\Gamma =-1$. The three panels show the histograms for the core extraction parameter sets \#3 (top), \#5 (middle), and \#9 (bottom). Each panel shows five histograms, corresponding to different degrees of crowding: the simulations with no overlaps (thick and solid black line), and the simulations where the mean separation to diameter ratio is 2.0 (dotted black line), 1.5 (thin and solid black line), 1.0 (dashed red line), and 0.5 (dotted blue line). The histograms are normalized to their maximum values.
              }
         \label{fig:ratio}
   \end{figure}
\begin{figure}
   \centering
   \includegraphics[bb=22 0 550 280, clip=true,width=\columnwidth]{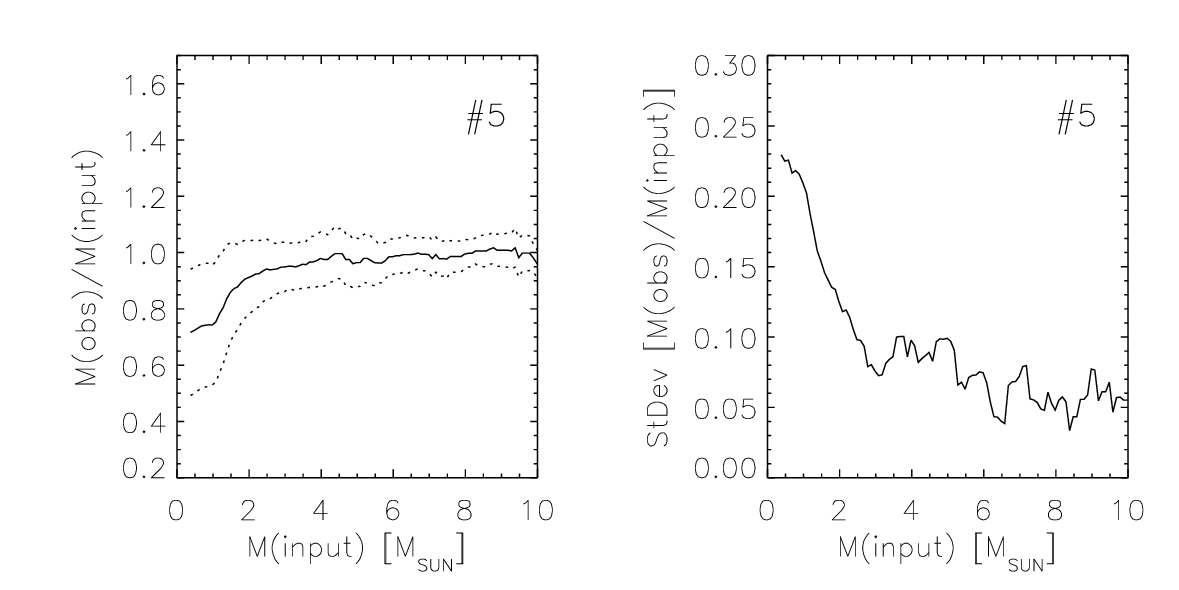} 
      \caption{Accuracy of the observed core masses as a function of the input mass of the cores in the simulation where $\Gamma = -1$ (contour level set \#5). \textbf{Left: }The mean of the $M_\mathrm{observed} / M_\mathrm{input}$ ratio. The standard deviation of the mean is shown with dotted lines. \textbf{Right: } The standard deviation of the $M_\mathrm{observed} / M_\mathrm{input}$ ratio.
              }
         \label{fig:accuracy}
   \end{figure}

The frequency distributions of the calculated $M_\mathrm{observed} / M_\mathrm{input}$ ratios for the simulations where $\Gamma =-1$ are shown in Fig. \ref{fig:ratio} with thick black lines. The three panels of the figure correspond to the contour levels \#3, \#5, and \#9 (see table \ref{tab:parameters}). The shapes of the distributions are relatively similar for different contour levels and close to Gaussians in all cases. The distributions have practically the same mean and median values of 0.85 and standard deviations of $0.25\dots 0.3$, being slightly larger for parameter sets \#3 and \#5 than for \#9. We note that the histograms presented in Fig. \ref{fig:ratio} are composed of all cores detected in those  simulations. This means that they are most representative for low mass cores that constitute the majority of the detected cores. The accuracy of the observed mass as a function of the input mass is illustrated in Fig. \ref{fig:accuracy} for one simulation (contour levels \#5). The figure shows both the $M_\mathrm{observed} / M_\mathrm{input}$ ratio and its standard deviation as a function of input mass. The standard deviation of the $M_\mathrm{observed} / M_\mathrm{input}$ ratio is 5-10 \% at $M_\mathrm{input} \gtrsim 2$ M$_\odot$. Below that mass, the standard deviation increases more rapidly, being 20-25 \% at $M_\mathrm{input} < 1 $ M$_\odot$. 

\subsection{The effect of crowding}


Because cores will not be isolated features in any real data, we need to determine the effect that core crowding has on the derived CMF. The cores that at least partially overlap may not necessarily be recognized as separate entities by the core extraction algorithm that is based only on 2-dimensional contouring. This may lead to an apparent absence of low-mass cores, and correspondingly to an overabundance of high-mass cores. In addition, the determination of the core masses suffers evidently from crowding, because in the contour-cutting algorithm each pixel is assigned uniquely to only one core, if any. As a consequence, if the crowding is severe enough the form of the true CMF can be totally lost. Our aim is to determine the fidelity limits at different degrees of crowding, and how crowded regions can be reliably studied with this method. 

\begin{figure*}
   \centering
   \includegraphics[width=1.95\columnwidth]{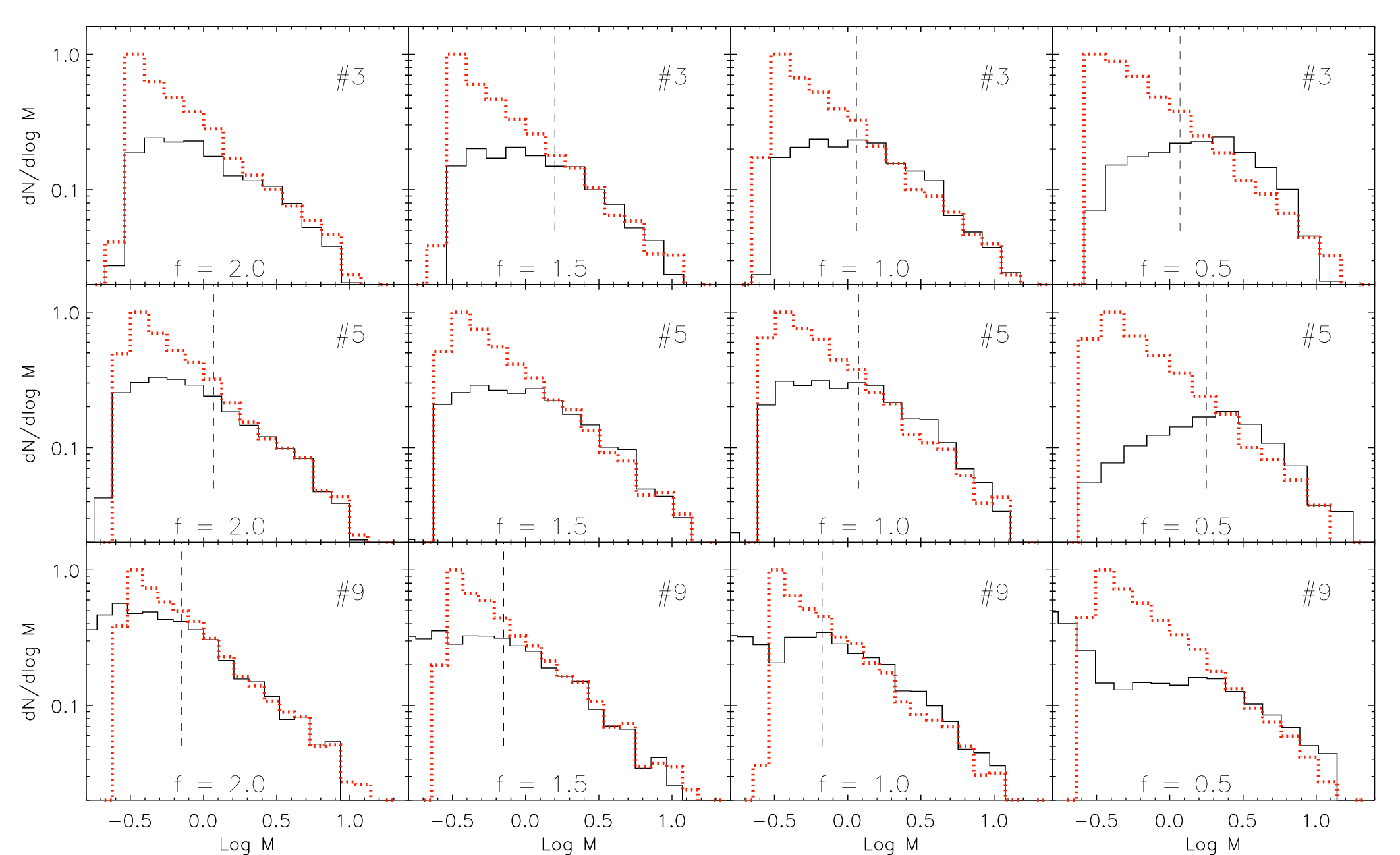}
      \caption{The CMFs derived for the simulations with different degrees of crowding. Each row shows the CMFs for four degrees of crowding, namely $f=2.0$, 1.5, 1.0, and 0.5. The three rows show the same CMFs derived using the core extraction parameter sets \#3 (top), \#5 (middle), and \#9 (bottom). The panels show the observed CMF with a black solid line and the input CMF with a red dotted line. The dashed vertical lines show the mass below which the derived CMF less than 90 \% complete. The histograms are normalized to the peak value of the input CMF. The slope of the input mass function in these simulations was $\Gamma =- 1$, and the range of input masses was $\log M = -0.55 \dots 1.2$. The input CMFs are always the same in each panel of the columns. However, the histograms are binned equally with the observed CMFs, and may therefore differ slightly from each other. 
                    }
         \label{fig:cmfs}
   \end{figure*}

To achieve this, we performed simulations where the ratio of mean separation to mean diameter was $f=2.0, 1.5, 1.0$, and 0.5. Figure \ref{fig:cmfs} shows the CMFs for these simulations in the case where the exponent of the input mass function is $\Gamma =-1$. The figure shows again the CMFs derived with the core extraction parameter sets \#3, \#5, and \#9 (see table \ref{tab:parameters}). 


There is a clear change in the shape of the observed CMFs at different degrees of crowding. In the most sparse simulations where $f=2$, the CMFs are well recovered above $M\gtrsim 1.5$ M$_\odot$ with parameter sets  \#1-\#4, above $M\gtrsim1.2$ M$_\odot$ with parameter sets  \#5-\#6, and above $M\gtrsim0.8$ M$_\odot$ with parameter sets  \#7-\#10. The simulations where $f=1.5$ yield very similar results. The fidelity limits with these degrees of crowding are actually close to the simulations where the cores did not overlap (Fig. \ref{fig:mfs_set_and_gamma}). Thus, the crowding seems to have negligible effect on the fidelity limit when $f\gtrsim 1.5$.

When the mean separation of the cores equals to the mean diameter of the cores ($f=1$), the observed CMFs begin to be more clearly altered. This is a rather logical level for the core crowding to become a severe issue for Gaussian shaped cores, because at this level the fraction of the mass of a core that is blended with neighboring cores becomes substantial compared to the original mass of the core. The effect of crowding is seen predominantly as an overabundance of cores at intermediate mass bins ($\log M \approx 0.5$), which results from the growing number of cores that are blended together. The fidelity limit indicated by the direct comparison of input and output mass bins are close to those where $f > 1.0$, but we emphasize that this limit is less meaningful when the crowding is this severe. 
The shape of the derived CMF above this limit is not accurately recovered when $f=1.0$. 

In the most crowded simulation with $f=0.5$ the observed CMFs recover the underlying mass function very inaccurately. The observed CMFs turn over at $M\approx 2.5$ M$_\odot$ regardless of the core extraction parameters, and there is a serious overabundance of cores above that mass. Clearly, if the crowding is this severe the observed CMF cannot be regarded to represent the underlying CMF.

The degree of core crowding is strongly related to the accuracy with which the core masses are recovered (Fig. \ref{fig:ratio}). 
In the simulations where $f=2.0$ (dotted black line) the standard deviation of the $M_\mathrm{obs} / M_\mathrm{input}$ distribution is almost similar to the case where the cores do not overlap (thick black line), but there is a more obvious tail in the distribution towards higher ratios. When the core crowding is more severe ($f=1.5$ and $f=1.0$) the distribution begins to deviate increasingly from the Gaussian shape. The standard deviation clearly increases and the tail towards higher ratios becomes more prominent. Because of this, also the mean and median values are higher. In the simulations with $f=1.0$, the median of the ratio is 1.0 and the standard deviation is 0.6. In the most crowded simulations with $f=0.5$ the distribution is clearly different from the previous cases, being exceedingly wide. The overlapping of cores makes the reliable determination of core masses practically impossible.

\section{Discussion}
\label{sec:discussion}


Near-infrared extinction data provide probably the most accurate and efficient way to trace the column density distribution of the most nearby molecular clouds over a large spatial and dynamical scale. Thus, $A_\mathrm{V}$ data can be exploited to study the CMF of dense cores. However, the derivation of the CMF based on extinction data alone can be hampered by difficulties in identifying the cores and measuring their masses and sizes from the 2-dimensional data. The accuracy at which the identification can be done depends undoubtedly on the adopted identification scheme, and on the details of the core population itself.

In this work we studied how accurately the CMF can be derived using extinction data and the core identification scheme used recently by ALL07 to derive the CMF of the Pipe nebula. Our results, presented in Section \S\ref{sec:results}, dictate that the CMF can indeed be determined reliably with this method, although some attention should be paid on selecting reasonable parameters for the core extraction algorithm. In our simulations, the identification performed with parameter sets \#5, \#7, and \#9 (see Table \ref{tab:parameters}) resulted to the most accurate reproduction of the input CMFs with the fidelity limits at $M=0.8$-1.5 M$_\odot$. Considering the CMF of the Pipe nebula derived by ALL07 this result confirms that the break of the power-law occuring at $M\approx 2$-3 M$_\odot$ is not produced by incompleteness or by bias in the core identification method. However, the simulations suggest significantly higher fidelity limits than estimated by ALL07, and hence introduce some restrictions to the use of method.


The most important factor hampering the identification method was found to be the mean separation of the cores within the cloud. By comparing the input and output CMFs, as well as input and output masses of individual cores, we concluded that the observed CMF recovers the underlying CMF adequately if the cores are on average smaller than their separations ($f \gtrsim 1$). This is not a totally unexpected result, considering that the identification method is based on contour line cutting and is therefore completely unable to deal with any kind of core overlaps. The statistical uncertainty in the masses of detected cores was found to be 25-30 \%, provided that $f \gtrsim 1.5$. However, the core masses are systematically underestimated by about 15 \% on average. This is due to the low column density outer-edges of the cores that remain undetected because the extraction algorithm artificially truncates the cores according to the lowest contour level. We note the 15 \% systematic error represents the mean ratio of the observed and input masses in the simulations (see also Fig. \ref{fig:accuracy}, left panel), and may depend on the specifics of the adopted core profile. 
 

In real clouds the $f$ ratio is indeed expected to be such that the core crowding may impair the use of the method. In the Pipe nebula, which is relatively quiescent and sparse, the ratio is $f\approx 2.0$. In complexes where star formation is more prominent, the cores are likely to be harbored in a substantially smaller volume than they are in the Pipe nebula. For example, the active central cluster of Rho Ophiuchi covers an area of about $1200$ arcmin$^2$, defined roughly as a region where $A_\mathrm{V}>10$ mags (according to the extinction map by Lombardi et al. \cite{lombardi08b}), and it contains $\sim 100$ cores (estimated from the sub-mm study of Stanke et al. \cite{stanke06}). We can calculate a first-order estimate for the core crowding by assuming that the region is a projection of a spherical volume inside which the cores are located. If 100 cores are distributed randomly into the volume, the sampling function of the projected nearest neighbor distribution has the mean of about $2'$. If the radii of the cores are similar to the cores of the Pipe nebula, i.e. $R \approx 5'$, the ratio $f \approx 0.4$ follows. If 200 cores are distributed into the volume, following the approximate number of stars in the cluster (Wilking et al. \cite{wilking05}), the ratio $f\approx 0.3$ follows. This implies that the method studied in this paper could not be employed in Rho Ophiuchi using similar dust column density data as was used for the Pipe nebula by ALL07, unless the cores are on average about 3 times smaller than the cores in the Pipe nebula. 


When working with data from real clouds the determination of the CMF is affected also by another important feature specific to the extinction data. Due to the strong and variable background extinction component, there is always confusion between real cores and projected spurious core-like structures of the background component. Even though the large-scale, smoothly varying component is filtered out by the wavelet decomposition, projected structures at the scale of real cores may pass the wavelet decomposition and be included in the cores only map. This is not a problem with high-mass cores, because they have large density contrasts with the local background. Therefore, having projected features mimicking large cores is very unlikely. In contrast, the local background may exceed the average extinction of low-mass cores by factors of several, making the occurrence of confusing structures more likely. It is not trivial to estimate the mass range where these false detections become important, but our experiences with the Pipe nebula suggest that measuring the CMF accurately below $M\lesssim 0.5$ M$_\odot$ would require a further analysis of this effect. As a consequence, when working with real data, one may be forced to compromise between extraction parameters yielding lower fidelity limits and parameters definitely eliminating the incorrect detections. The confusion may also be solved by imposing some more sophisticated prerequisites, such as detection of any associated dense gas traced by molecular lines or multiwavelength continuum observations, for a column density peak to be counted as a real core. 


Finally, we comment on the applicability of our simulations. First, the background extinction field used in the simulations was always the same, i.e. the wavelet subtracted extinction map of the Pipe nebula. This field has its specific features, characterized by, for example, the frequency distribution of the extinction pixel values. Although we expect that the results of the simulations are generally applicable to other clouds, some caution is appropriate especially if the characteristics of the extended extinction component are rigorously different from those of the Pipe nebula. Second, in the simulations all the massive cores are single, well defined objects. In a real cloud, cores whose masses are significantly in excess of the Jeans' or Bonnor-Ebert critical mass are likely to exhibit substructure and may be fragmented.  Under these circumstances it is not clear how to incorporate such cores into a CMF.  Should such a core be counted as a single massive core or broken up and counted as a number of smaller mass cores? The strategy one adopts to deal with this issue will necessarily influence the choice of input parameters for the core identification algorithm. For example, low detection thresholds and small contour level spacing (e.g., contour levels \# 9) will tend to break up massive cores with substructure into individual components, while high detection thresholds and larger intervals (e.g., contour levels \#3) will suppress substructure and identify such substructured cores as single massive objects. 

Third, even though the shape of the observed CMF above the fidelity limit suggested by the simulations is not affected by the identification method, the shape is still subject to the normal counting errors. In the traditional scheme where a power-law is fitted to the logarithmically binned differential CMF, it would require $N>500$ cores to reduce the actual error of the fitted power-law index below $0.1$ (e.g. Rosolowsky \cite{rosolowsky05}). In our simulations the counting errors of the CMFs were deliberately reduced, because our purpose was to examine uncertainties originating from the identification method only.

\section{Conclusions}
\label{sec:conclusions}

We performed simulations where we examined the recoverability of core masses and the accuracy and completeness limit of the CMF derived for a molecular cloud using dust column density data. In particular, we studied a method where the cores are extracted from the background extinction using a wavelet decomposition and a 2-dimensional thresholding routine, \textsf{clfind2d}. In the simulations we generated a variable field of background extinction with populations of artificial cores embedded in it. We extracted the cores and compared the mass function constructed from them to the input mass function. We also studied how the degree of core crowding affects the fidelity limits and the accuracy of the derived core masses. The main conclusions of our work are as follows.

\begin{enumerate}

      \item The selection of parameters for the core extraction algorithm has an effect on the accuracy and on the completeness limit of the derived CMF. However, a more important factor is the crowding of the cores. We find that the CMF and the core masses can be reliably determined if the ratio of the mean separation to the mean diameter ($f$) of the cores is larger than one, i.e. if the cores are on average smaller than their separations. 
       
      \item Provided that $f>1$, the derived CMF resembles the underlying mass function above $M>0.8$-1.5 M$_\odot$, depending on the selection of the core extraction parameters. In the simulations the lowest fidelity limit was achieved with a relatively low threshold level and narrow level spacings, namely $A_\mathrm{V}^\mathrm{th} = 1$ or $2 \sigma^\mathrm{rms}$, and $\delta A_\mathrm{V} = 3 \sigma^\mathrm{rms}$. Choosing larger threshold and wider level spacings results in higher fidelity limits. If $f<1$, the observed CMF may not represent the underlying mass function, and thus the CMF cannot be reliably measured using the method studied in this paper.
      
      \item The masses of individual cores are recovered with a typical uncertainty of 25-30 \%, provided that $f\gtrsim 1.5$. In the range $1.5 > f > 1.0$ the uncertainty is about 60 \%. If $f < 1$ the mass determination of individual cores is very uncertain.
            
\end{enumerate}





\begin{thebibliography}{}

\bibitem[2007]{alves07} Alves, J., Lombardi, M., \& Lada, C.~J.\ 2007, \aap, 462, L17, ALL07

\bibitem[1978]{bohlin78} Bohlin, R.~C., Savage, B.~D., \& Drake, J.~F.\ 1978, \apj, 224, 132 

\bibitem[2008]{enoch08} Enoch, M.~L., Evans, N.~J., II, Sargent, A.~I., et al.\ 2008, \apj, 684, 1240 


\bibitem[2006]{johnstonebally06} Johnstone, D., \& Bally, J.\ 2006, \apj, 653, 383 

\bibitem[2006]{johnstone06} Johnstone, D., Matthews, H., \& Mitchell, G.~F.\ 2006, \apj, 639, 259 

\bibitem[2001]{johnstone01} Johnstone, D., Fich, M., Mitchell, G.~F., \& Moriarty-Schieven, G.\ 2001, \apj, 559, 307 

\bibitem[2000]{johnstone00} Johnstone, D., Wilson, C.~D., Moriarty-Schieven, G., et al.\ 2000, \apj, 545, 327 

\bibitem[2002]{kroupa02} Kroupa, P.\ 2002, Science, 295, 82 

\bibitem[2008]{lada08} Lada, C.~J., Muench, A.~A., Rathborne, J., Alves, J.~F., \& Lombardi, M.\ 2008, \apj, 672, 410 

\bibitem[2008]{lombardi08} Lombardi, M.\ 2008, arXiv:0809.3383v1 [astro-ph]

\bibitem[2008b]{lombardi08b} Lombardi, M., Lada, C.~J., \& Alves, J.\ 2008b, \aap, 480, 785 

\bibitem[2006]{lombardi06} Lombardi, M., Alves, J., \& Lada, C.~J.\ 2006, \aap, 454, 781, LAL06

\bibitem[2001]{lombardi01} Lombardi, M., \& Alves, J.\ 2001, A\&A, 377, 1023 

\bibitem[1990]{mathis90} Mathis, J.~S.\ 1990, \araa, 28, 37 

\bibitem[2001]{motte01} Motte, F., Andr{\'e}, P., Ward-Thompson, D., \& Bontemps, S.\ 2001, \aap, 372, L41 

\bibitem[1998]{motte98} Motte, F., Andre, P., \& Neri, R.\ 1998, \aap, 336, 150 

\bibitem[2007]{nutter07} Nutter, D., \& Ward-Thompson, D.\ 2007, \mnras, 374, 1413 

\bibitem[2006]{reid06} Reid, M.~A., \& Wilson, C.~D.\ 2006a, \apj, 650, 970 

\bibitem[2006]{reid06a} Reid, M.~A., \& Wilson, C.~D.\ 2006b, \apj, 644, 990 

\bibitem[2005]{rosolowsky05} Rosolowsky, E.\ 2005, \pasp, 117, 1403 

\bibitem[2006]{stanke06} Stanke, T., Smith, M.~D., Gredel, R., \& Khanzadyan, T.\ 2006, \aap, 447, 609

\bibitem[1955]{salpeter55} Salpeter, E.~E.\ 1955, \apj, 121, 161

\bibitem[2008]{simpson08} Simpson, R.~J., Nutter, D., \& Ward-Thompson, D.\ 2008, ArXiv e-prints, 807, arXiv:0807.4382 

\bibitem[1998]{testi98} Testi, L., \& Sargent, A.~I.\ 1998, \apjl, 508, L91 


\bibitem[2006]{young06} Young, K.~E., Enoch, M.~L., Evans, N.~J.~II, et al.\ 2006, \apj, 644, 326 

\bibitem[2005]{wilking05} Wilking, B.~A., Meyer, M.~R., Robinson, J.~G., \& Greene, T.~P.\ 2005, \aj, 130, 1733 

\bibitem[1994]{williams94} Williams, J.~P., de Geus, E.~J., \& Blitz, L.\ 1994, \apj, 428, 693 

\end{thebibliography}
\end{document}